\documentclass[aps,showpacs,preprint,superscriptaddress]{revtex4}
\usepackage[caption = false]{subfig}
\usepackage[final]{graphicx}
\usepackage{float}
\usepackage{bm}
\usepackage{amsmath}
\usepackage{txfonts}
\usepackage{epstopdf}
\usepackage{natbib}
\begin{document}

\title{Electron-positron Pair Creation by Counterpropagating Laser Pulses: Role of Carrier Envelope phase}
\author{Chitradip Banerjee\footnote{Corresponding author.\\E-mail address: cbanerjee@rrcat.gov.in (C. Banerjee).}}
\author{Manoranjan P. Singh}
\affiliation{Theory and Simulations Lab, HRDS, Raja Ramanna Centre for Advanced
Technology,  Indore-452013, India}
\affiliation{Homi Bhaba National Institute, Training School Complex, Anushakti Nagar, Mumbai 400094, India}

\begin{abstract}
The effect of carrier envelope phase (CEP) on the spatio-temporal distribution of the electron-positron pairs created by untraintense counterpropagating femtosecond laser pulses is studied. When the laser pulses are linearly polarized, the temporal distribution of the pairs is found to be sensitive to CEP. On the hand, it is found to be largely insensitive to CEP for circularly polarized pulses. 
\end{abstract}
\maketitle
\section{Introduction}
Particle-antiparticle (here $e^+e^-$) pair production is of fundamental interest for studying the non-linear processes in the presence of strong electric field interacting with the vacuum state in the realm of quantum electrodynamics (QED) \cite{RevModPhys.84.1177DiPiazza}. The strength of electric field $E_S$ which is needed to have a real $e^+e^-$ pair is called the characteristic field of QED. It is also known as the Schwinger limit and its value is  $1.32\times 10^{16}\textnormal V/cm$ \cite{Sauter}. The process in which the $e^+e^-$ pairs are generated from vacuum in the presence of such  electric fields in the non-perturbative regime, is known as the Schwinger mechanism \cite{PhysRev.82.664}. It has been well studied for the different kind of fields such as homogeneous electric field constant in time, time-varying but spatially homogeneous electric fields \cite{Sauter,PhysRev.82.664,NarozhnyiNikishov1970,PhysRevD.2.1191Itzykson} and 
the  fields inhomogeneous in both space and time \cite{PhysRevE.69.036408,PhysRevLett.104.220404}. Several experimental projects are under way to envisage the nonlinear QED effects with soon-coming laser facilities like Extreme Light Infrastructure (ELI) and High Power laser Energy Research (HiPER) \cite{NaturePublishingGroupDunne}. Of late, these studies have received a  renewed attention due to the radical advancement in the field of laser technology \cite{PhysRevSTAB.5.031301} leading to enormous increase in the achievable field strengths even though the present day available laser intensity $I$ ($\approx 10^{22}\textnormal W/cm^2$) is still far below the critical intensity $I_{cr}=\frac{c}{4\pi}E_S^2 \approx 4.6\times 10^{29}\textnormal W/cm^2$. 

For the Schwinger mechanism, two basic criteria need to be fulfilled by the electromagnetic(EM) field. First, the EM field should have non zero Lorentz invariants and second, the peak electric field strength should be close to the critical field. A simple way to meet these criteria is to use focused ultrashort  and ultraintense laser beam(s). The focused laser beams can be described by the various field models such as  Narozhny-Fofanov (NF) field model for weakly focused pulse \cite{Narozhny2000}, tightly focused field model  \cite{FedotovA.M.tightfocus,PhysRevSTAB_salamin} and optimally focused field model \cite{PhysRevA.86.2012Gonoskov,PhysRevLett.111.060404Gonoskov}. Use of two or many counterpropagating laser beams have been shown to reduce the intensity threshold much below the critical intensity  \cite{bula,PhysRevLett.104.220404}. In these field configurations the magnetic field vanishes in the focal region and thereby enhancing the pair production rate \cite{PhysRevA2012Su,PhysRevLett.109.253202Su}. Furthermore, because of the formation of the standing wave pattern in the focal plane the peak field strength of the electric field increases. Consequently several theoretical studies have explored various aspects of pair production using the counterpropagating beam configuration of focused laser beams such as, enhancement of the production rate \cite{bula,PhysRevLett.104.220404,FedotovA.M.tightfocus}, momentum distribution of the created particles \cite{AbdukerimCarrier,Orthaber201180}, dynamically assisted Schwinger mechanism \cite{PhysRevD.82Dumlu,NurimanAbdukerimchinPhysB,Sitiwaldi2017174,PhysRevD.88.045028Kohlfurst,PhysRevLett.102.080402Ruf,PhysRevD.82.105026Hebenstreit,PhysRevLett.106.035001Nerush}, and spin-polarization state of the created particles on the laser field polarization \cite{PhysRevD.91.125026Wollert}.

The ultrashort laser pulse can be described by the pulse envelope function and the carrier envelope phase (CEP) which is basically the phase difference between the carrier wave and the envelope of the pulse profile \cite{Dipiazza_CEP,RevModPhys_CEP1_nonlinear,RevModPhys_CEP}. CEP of ultrashort pulses, particularly with a few cycles, can have significant bearing on the QED processes. In fact, it can be used to determine CEP as reported in Ref. \cite{Dipiazza_CEP} wherein the angular distribution of photons emitted by an electron via multiphoton Compton scattering due to an intense laser pulse has been shown to be directly related to the CEP of the laser pulse. The effect of CEP on the momentum distribution of the produced pairs has also been extensively explored \cite{PhysRevLett_CEP2009,PhysRevD.82Dumlu,AbdukerimCarrier,NurimanAbdukerimchinPhysB,Orthaber201180,Sitiwaldi2017174}.

In this paper we investigate the effects of CEP on the invariant field distribution and the differential pair yields for counterpropagating laser pulses made up of e-waves \cite{Narozhny2000,2016arXiv160601367B}. It is found that CEP does not have any appreciable effect on the invariant electric and magnetic fields in the focal region when the laser pulses are circularly polarized, see in \textbf{Appendix B}. For this reason we present results for linearly polarized laser pulses for which CEP is found to affect the temporal distribution of the invariant fields and the created pairs thereof.

The paper is organized as follows. In Sec.\ref{Theory} we discuss the EM field configurations for linearly e-waves laser pulses. The dependence of the CEP on the energy flow from the focal volume is studied. The distributions of the invariant fields and the created particles is discussed in \ref{result}. Finally we  conclude in Sec.\ref{concl}. The technical details of the analytical expressions for the EM field and field invariants are given in \textbf{Appendices A} and \textbf{B}.
\section{Theoretical Method }\label{Theory}
\subsection{The field model}\label{field_theory}
We consider the EM field structure of two linearly polarized counterpropagating focused laser pulses. Using NF field model \cite{Narozhny2000} the expressions of the laser pulses propagating in $z$ and $-z$ directions and having their focal region centred about the origin for the case where the pulses are made up of two e-waves are given as (detailed calculations have been shown in \textbf{Appendix A}): 
\begin{equation}\label{E_e_vec}
\begin{split}
\textbf{E}^e = 2E_0g\frac{e^{-\xi^2/{(1+4\chi^2)}}}{1+4\chi^2}\sin(\omega t+\tilde{\varphi})\left[\hat{\textbf{e}}_x\left\{\cos(\omega z/c-2\psi)-\frac{2\xi^2}{(1+4\chi^2)^{1/2}}\sin^2{\phi}\cos(\omega z/c-3\psi)\right\}\right.\\\left.+\hat{\textbf{e}}_y\frac{\xi^2}{(1+4\chi^2)^{1/2}}\sin{2\phi}\cos(\omega z/c-3\psi)\right],
\end{split}
\end{equation}
and 
 \begin{equation}\label{H_e_vec}
\begin{split}
\textbf{H}^e \approx -2E_0g\frac{e^{-\xi^2/{(1+4\chi^2)}}}{1+4\chi^2}\left[\cos(\omega t+\tilde{\varphi})\Bigg(\hat{\textbf{e}}_x\frac{\xi^2}{(1+4\chi^2)^{1/2}}\sin{2\phi}\sin(\omega z/c-3\psi)\right.\\-\left.\hat{\textbf{e}}_y\left\{\sin(\omega z/c-2\psi)-\frac{2\xi^2}{(1+4\chi^2)^{1/2}}\sin^2{\phi}\sin(\omega z/c-3\psi)\right\}\Bigg)\right.\\+\left.4\xi\Delta\frac{\sin(\omega t+\tilde{\varphi})}{(1+4\chi^2)^{1/2}}\sin{\phi}\sin(\omega z/c-3\psi)\hat{\textbf{e}}_z\right].
\end{split}
\end{equation}
Here $E_0$ is the peak electric field strength of the laser beams, $\omega$ is the corresponding central frequency, $\lambda$ is the wavelength, $\Delta$ is the focusing or spatial inhomogeneity parameter, $R$ is the focusing radius, $L$ is the Rayleigh length; and $\xi = \rho/R,\:  \chi = z/L,\: 
  \rho = \sqrt{x^2+y^2},\:  \exp(i\phi) = (x+iy)/\rho,\:
  \Delta = c/{\omega R} = \lambda/{2\pi R}, \: L = R/\Delta,\:\textnormal{and}\: \exp(i\psi) = (1+2i\chi)/\sqrt{1+4\chi^2}$.
Here superscript e refers to the focused EM field configuration in which the electric field is transverse to the propagation direction \cite{Narozhny2000}.
In Eqs.(\ref{E_e_vec}-\ref{H_e_vec}), $g$ is the temporal envelope function to account for the finite pulse width and $\tilde{\varphi}$ is the corresponding CEP. Though there can be various functional forms of $g$ consistent with the condition that $g(0) = 1$ and $g$ should decrease very fast at the periphery of the focal pulse for $|\varphi|\gg\omega\tau$, we take $g = \exp(-4(t^2/\tau^2+z^2/{c^2\tau^2}))$ for all the calculations presented here \cite{Narozhny2000, bula, 2016arXiv160601367B}. Here $\varphi = \omega(t-z/c)$ is defined as the dynamic phase of the laser EM wave. The strength of the resultant EM field due to the counterpropagating laser pulses in the focal region ( $|\chi|<1, \xi <1$)  can be approximated as
\begin{equation}\label{E_e_mag}
|\textbf{E}^e| \approx\\ 2E_0g\frac{e^{-\xi^2/{(1+4\chi^2)}}}{1+4\chi^2}|\sin(\omega t+\tilde{\varphi})|\left|\cos(\omega z/c-2\psi)\right|\left[ 1-\frac{2\xi^2}{(1+4\chi^2)^{1/2}}\sin^2{\phi}\right],  
\end{equation}
and 
\begin{equation}\label{H_e_mag}
|\textbf{H}^e| \approx\\ 2E_0g\frac{e^{-\xi^2/{(1+4\chi^2)}}}{1+4\chi^2}|\cos(\omega t+\tilde{\varphi})|\left|\sin(\omega z/c-2\psi)\right|\left[ 1-\frac{2\xi^2}{(1+4\chi^2)^{1/2}}\sin^2{\phi}\right].  
\end{equation}
Since the pair creation process is solely governed by the invariant EM fields so it is worthwhile to calculate the EM field invariants. The expressions for the two Lorentz invariants $\mathcal{F}^e, \: \mathcal{G}^e$ of the EM field given by Eqs.(\ref{E_e_vec}-\ref{H_e_vec}) are	
\begin{equation}\label{F_e}
\mathcal{F}^e = \frac{1}{2} ({\textbf{E}^e}^2-{\textbf{H}^e}^2 )\approx\\ 2E_0^2g^2\frac{e^{-2\xi^2/{(1+4\chi^2)}}}{(1+4\chi^2)^2}\left\{\sin^2(\omega t+\tilde{\varphi})-\sin^2(\omega z/c-2\psi)\right\}\left[ 1-\frac{4\xi^2}{(1+4\chi^2)^{1/2}}\sin^2{\phi}\right],  
\end{equation}
and
\begin{equation}\label{G_e}
\mathcal{G}^e = \textbf{E}^e\cdot\textbf{H}^e\approx -4E_0^2g^2\xi^2\chi\frac{e^{-2\xi^2/{(1+4\chi^2)}}}{(1+4\chi^2)^{5/2}}\sin{2(\omega t+\tilde{\varphi})}\sin{2\phi}\left[ 1+3\xi^2\right].  
\end{equation}
 For e-wave beam configuration the reduced invariant electric and magnetic fields are defined as \cite{PhysRev.82.664,2016arXiv160601367B}: 
\begin{equation}\label{invariant_field}
\begin{split}
\epsilon^e = \frac{1}{E_S}\sqrt{\sqrt{{\mathcal{F}^e}^2+{\mathcal{G}^e}^2}+\mathcal{F}^e},\:\:\:\:\:
\eta^e = \frac{1}{E_S}\sqrt{\sqrt{{\mathcal{F}^e}^2+{\mathcal{G}^e}^2}-\mathcal{F}^e}.
\end{split}
\end{equation}
The value of $\mathcal{G}^e$ is negligibly small in the focal region. It has maximum in the peripheral region $\xi =0.75$, $\chi = \pm 0.25$ for $t = 0.001\tau$, $\phi = \pi/4$, and $\tilde{\varphi} = \pi/2$. Still this maximum value is $0.01$ times less than that of $\mathcal{F}^e$ at the space-time position. In the approximation where $\mathcal{G}^e$ can be neglected the sign of $\mathcal{F}^e$ (which is given by whether $\sin^2(\omega t+\tilde{\varphi})-\sin^2(\omega z/c-2\psi)$ is positive or negative) gives two non-trivial situations. If $\sin^2(\omega t+\tilde{\varphi}) > \sin^2(\omega z/c-2\psi)$, then $\mathcal{F}^e$ is positive and consequently $\epsilon^e$  survives and $\eta^e$ vanishes. This gives rise to what is known as electric regime \cite{ElectricregimeFedotov}: 
\begin{equation}\label{epsilon1_e}
\epsilon^e \approx 2E_0g\frac{e^{-\xi^2/{(1+4\chi^2)}}}{1+4\chi^2}\left\{\sin^2(\omega t+\tilde{\varphi})-\sin^2(\omega z/c-2\psi)\right\}^{1/2}\left[ 1-\frac{2\xi^2}{(1+4\chi^2)^{1/2}}\sin^2{\phi}\right],\:\:\textnormal{and}\:\:\eta^e \approx 0.  
\end{equation}
Similarly one has magnetic regime ($\mathcal{F}^e$ is negative) for the case when $\sin^2(\omega t+\tilde{\varphi}) < \sin^2(\omega z/c-2\psi)$ \cite{ElectricregimeFedotov}. $\epsilon^e$ vanishes and $\eta^e$ survives: 
\begin{equation}\label{epsilon2_e}
\epsilon^e \approx 0,\:\:\textnormal{and}\:\:\eta^e \approx\\ 2E_0g\frac{e^{-\xi^2/{(1+4\chi^2)}}}{1+4\chi^2}\left\{\sin^2(\omega z/c-2\psi)-\sin^2(\omega t+\tilde{\varphi})\right\}^{1/2}\left[ 1-\frac{2\xi^2}{(1+4\chi^2)^{1/2}}\sin^2{\phi}\right].  
\end{equation}
Recalling that $\epsilon^e$ and $\eta^e$ are the electric and magnetic fields strengths in the frame where they are parallel, it can be easily seen that expressions of the electric and magnetic fields in both the frames are not identical as seen in the Eqs.(\ref{E_e_mag},\ref{epsilon1_e}, \ref{epsilon2_e}, \ref{H_e_mag}). This may be contrasted with the corresponding expressions for circularly polarized laser beams where fields in both the frames are nearly identical in the focal region arising because of resultant lab frame electric and magnetic fields of the counterpropagating beams being nearly parallel in the focal region \cite{2016arXiv160601367B}. Since electric and magnetic fields in this case are not parallel, there will be flow of energy from the focal region governed by the Poynting vector ($\textbf{S}^e$). The $x$, $y$, and $z$ components of $\textbf{S}^e$ are given as
\begin{equation}\label{S_ex}
S^e_x \approx 16E_0^2g^2\xi^3\Delta\frac{e^{-2\xi^2/{(1+4\chi^2)}}}{(1+4\chi^2)^3}\sin^2(\omega t+\tilde{\varphi})\sin^2{\phi}\cos{\phi}\sin{2(\omega z/c-3\psi)},
\end{equation} 
\begin{equation}\label{S_ey}
\begin{split}
S^e_y \approx -16E_0^2g^2\xi\Delta\frac{e^{-2\xi^2/{(1+4\chi^2)}}}{(1+4\chi^2)^{5/2}}\sin^2(\omega t+\tilde{\varphi})\sin{\phi}\sin(\omega z/c-3\psi)\Bigg[\cos{2(\omega z/c-2\psi)}\\-\frac{2\xi^2}{(1+4\chi^2)^{1/2}}\sin^2{\phi}\cos(\omega z/c-3\psi)\Bigg],
\end{split}
\end{equation} 
and 
\begin{equation}\label{S_ez}
S^e_z \approx -E_0^2g^2\frac{e^{-2\xi^2/{(1+4\chi^2)}}}{(1+4\chi^2)^{2}}\sin{2(\omega t+\tilde{\varphi})}\left[\sin{2(\omega z/c-2\psi)}-\frac{4\xi^2}{(1+4\chi^2)^{1/2}}\cos^2{\phi}\sin(2\omega z/c-5\psi)\right].
\end{equation}

In Eqs.(\ref{S_ex},\ref{S_ey},\ref{S_ez}), the Cartesian components of $\textbf{S}^e$ show that the energy flow in $x$ and $y$ directions are much smaller than that in the $z$-direction. The oscillatory nature of $z$-component of Poynting vector $S_z^e$ leads to instantaneous energy flow whereas the average energy flow is zero. 

For e-linearly polarized counterpropagating laser pulses (e-LPCLP) beam discussed above, an additional control over the pair production mechanism can be achieved by tuning CEP with respect to the dynamic phase $\varphi$ of the laser pulses. In particular, if $\sin^2(\omega t+\tilde{\varphi}) < \sin^2(\omega z/c-2\psi)$ then EM field energy will remain confined within this region as a standing wave without any loss due to the $e^+e^-$ pair production. 
 
As discussed in \textbf{Appendix B}, for the circularly polarized counterpropagating laser beams CEP negligibly affect the field strengths and the invariants in the focal region. The above phase relationship does not maintain in  e-circularly polarized counterpropagating laser pulses (e-CPCLP) and EM field goes to self-attenuation by the generation of $e^+e^-$ near the critical field strength which we discuss in \textbf{Appendix B}. Such unavoidable energy loss of CPLP can be circumvented by using LPCLP.
\section{Results and discussions}\label{result}
\subsection{Field distribution}\label{field_dist}
The spatio-temporal distribution of $\epsilon^e$ for e-LPCLP as given in \ref{field_theory} is discussed here for different values of CEP. The space-time variables are scaled by the laser parameters such as: time is scaled by the pulse duration $\tau$; longitudinal variable $z$ is scaled by the Rayleigh length $L$; and the transverse variables $x$, $y$ are by the focusing radius $R$ of the laser beam.

\begin{figure}
\begin{center}
\subfloat[]{\includegraphics[width = 2.4in]{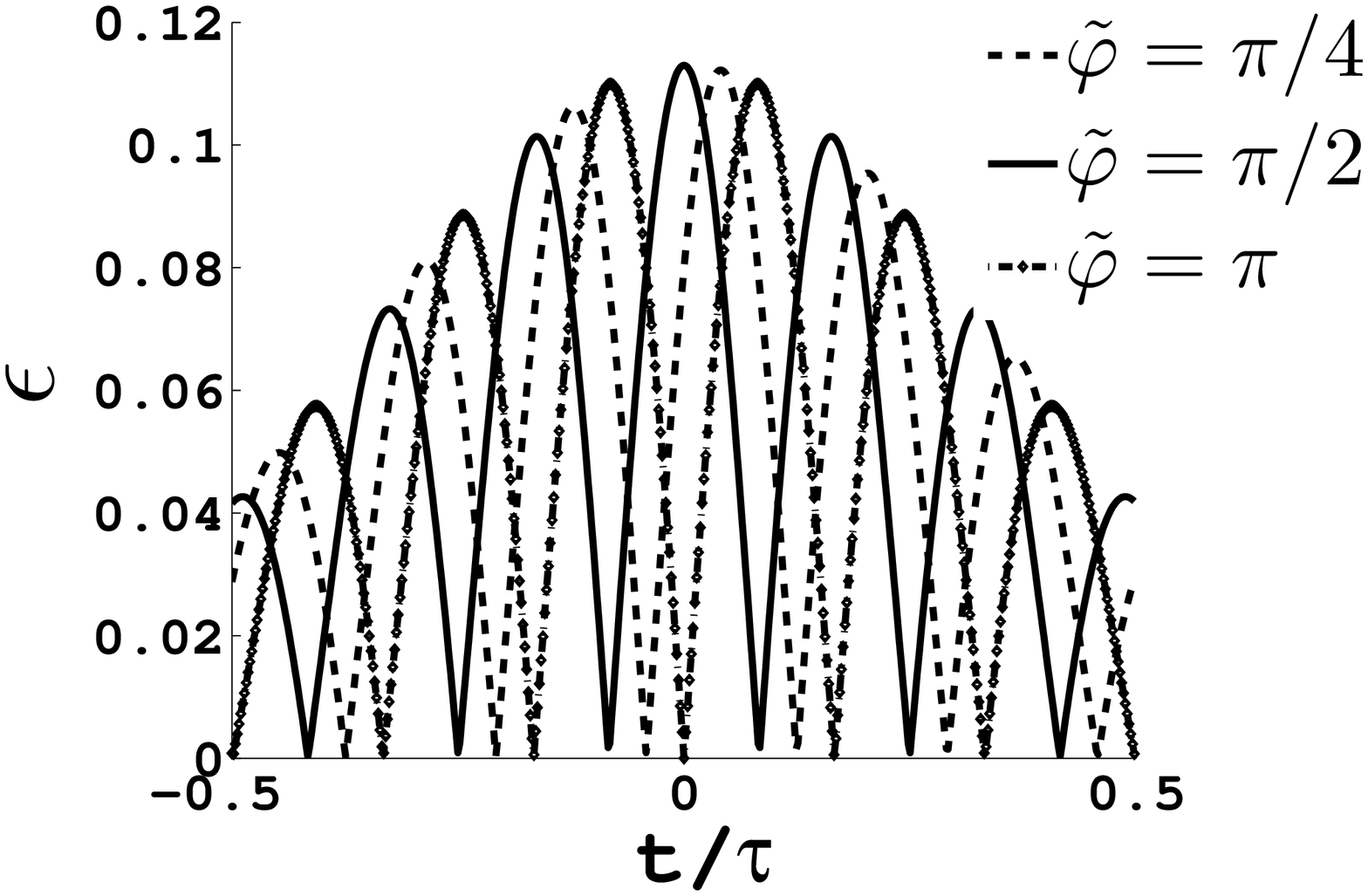}}
\subfloat[]{\includegraphics[width = 2.4in]{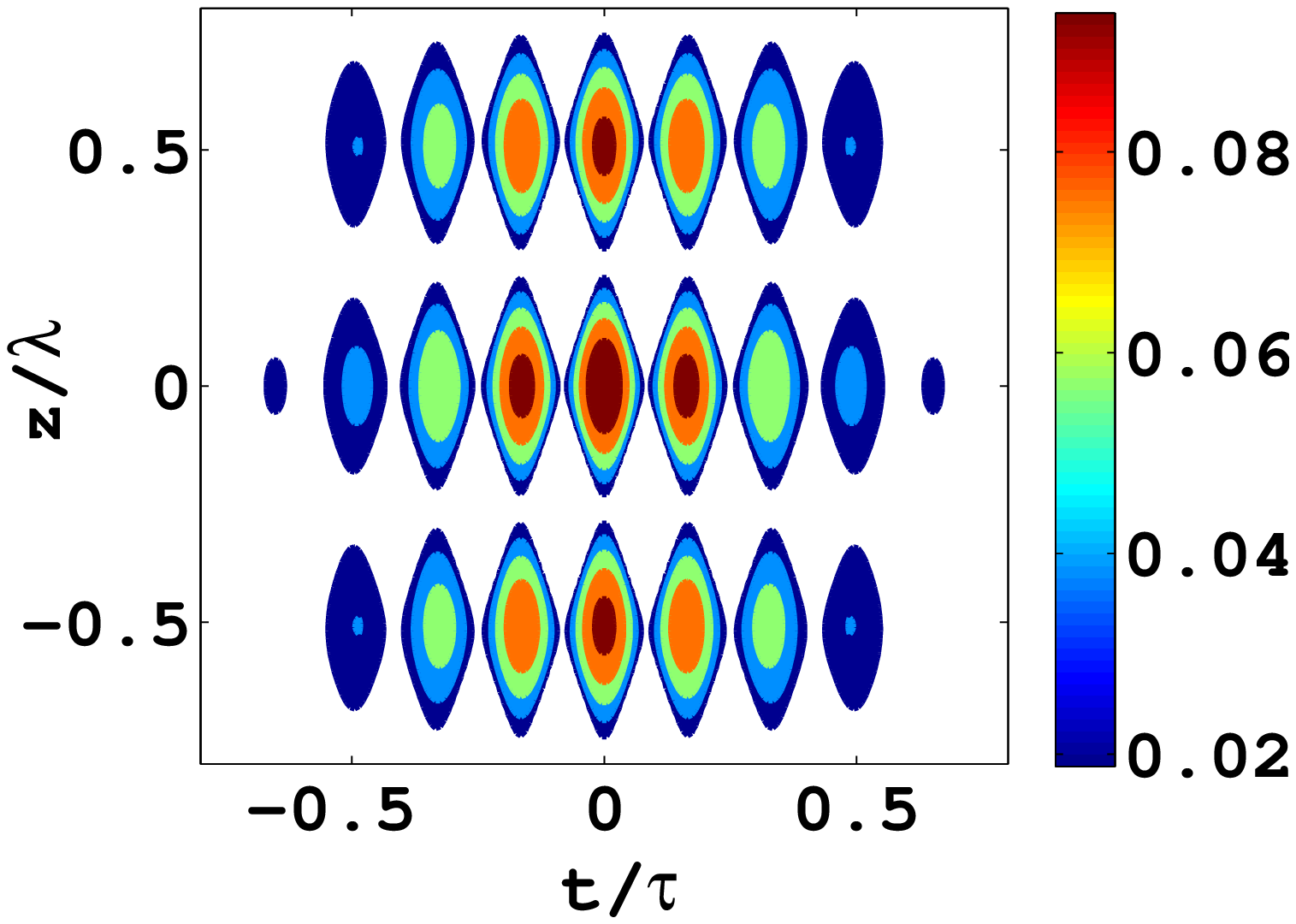}}\\ 
\subfloat[]{\includegraphics[width = 2.4in]{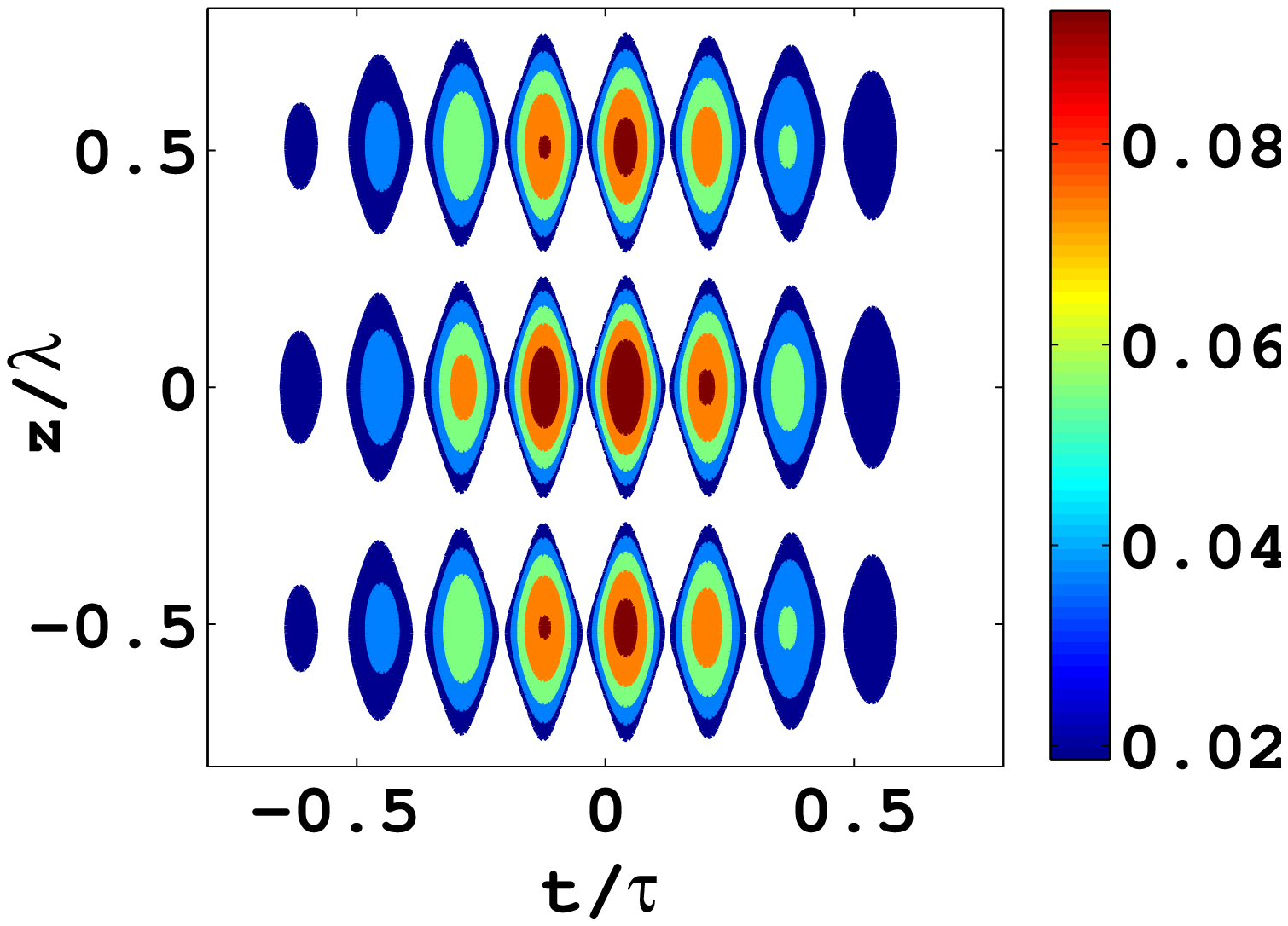}}
\subfloat[]{\includegraphics[width = 2.4in]{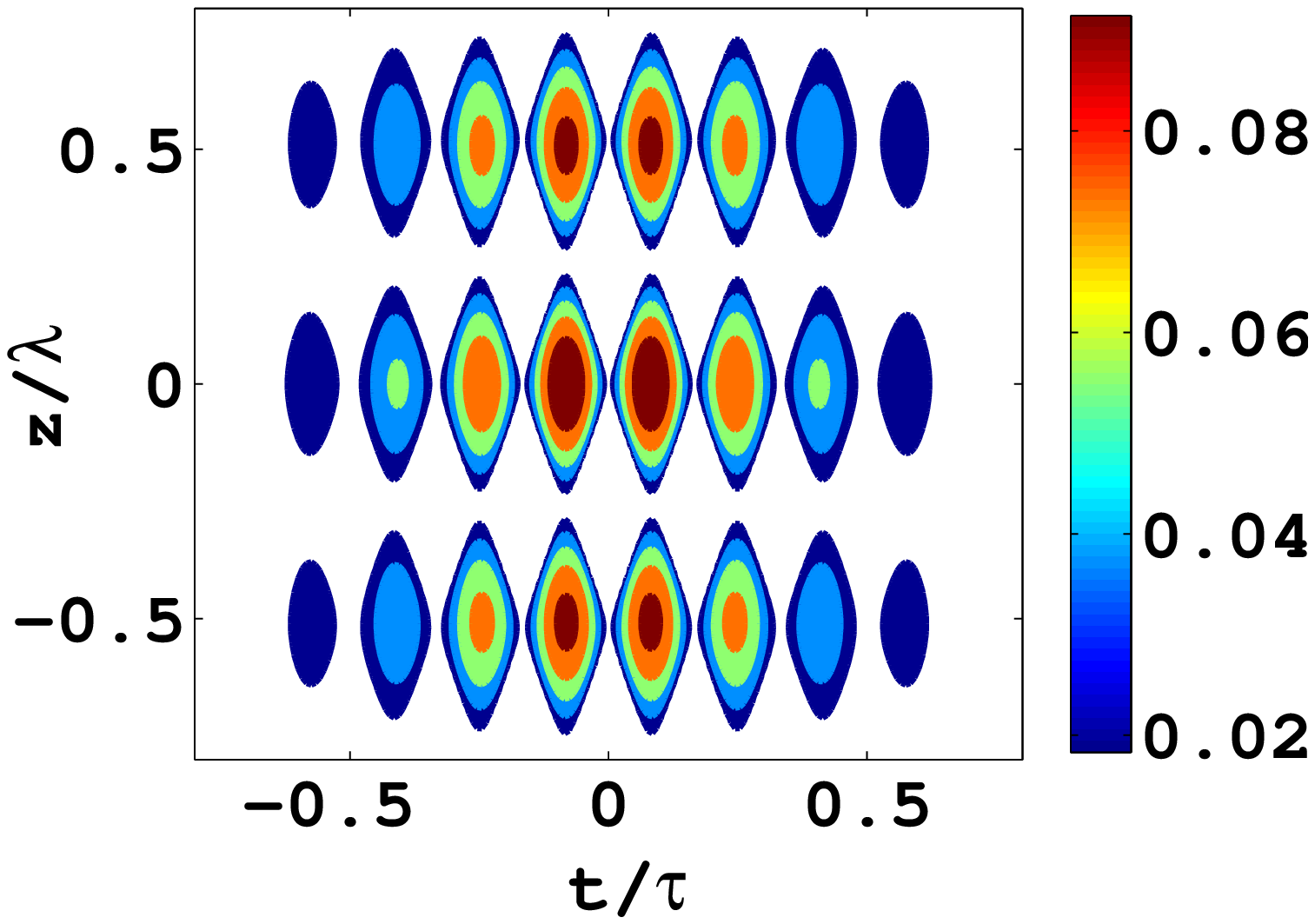}} 
\caption{The temporal evolution of $\epsilon^e$ and its contour plots in $zt$-plane in the scaled variables for different values of CEP, showing the locations of the peak field positions. Top left panel shows the temporal evolution of $\epsilon^e$ for $\tilde{\varphi} = 0$, $\pi/4$, and $\pi/2$. In the top right to bottom right panels the contour plots for $\tilde{\varphi} = \pi/2$, $3\pi/4$, and $\pi$ in the $zt$-plane for $x = y =0$ are given. The other field parameters are $E_0 = 0.0565$, $\Delta = 0.1$, $\tau = 10fs$, and $\lambda = 1\mu m$. The adjacent colour bars are showing the normalized field strength at the field peak positions.}
\label{epsilon_z_t}
\end{center}
\end{figure}

Fig.\ref{epsilon_z_t}(a) shows the distributions of $\epsilon^e$ with $t$ for CEP  $\tilde{\varphi} = 0$,  $\pi/4$, and  $\pi/2$ at the focal point. The invariant field shows oscillatory behaviour inside the pulse envelope due to the interference of counterpropagating pulses in the temporal domain. For $\tilde{\varphi} = \pi/2$ there is a central peak accompanied by the smaller peaks symmetrically placed on the either side of the central peak in the temporal profile of the invariant electric field. As the value of $\tilde{\varphi}$ is reduced to $\pi/4$, the temporal profile of $\epsilon^e$ shifts to the right, i.e. towards the leading part of the laser pulse. Moreover the profile becomes asymmetric in time. The peaks in the leading part of the laser pulse are smaller and the ones in the trailing part are large compared to those for $\tilde{\varphi} = \pi/2$. For $\tilde{\varphi} = 0$, the temporal profile is again symmetric. However, it has a minimum at the centre of the laser pulse and has two major maxima on either side of the centre.
The reduced magnetic field $\eta^e$ vanishes completely in this case (data not shown) as $\mathcal{G}^e$ is identically equal to zero and $\mathcal{F}^e$ is positive for $z = 0$. For $z\neq 0$ and $x = y = 0$ (where $\mathcal{G}^e = 0$),
depending on the sign of $\mathcal{F}^e$, we have a mesh like structure in the $z-t$ plane where some regions belong to the electric regime and other to the magnetic regime. 
In Fig.\ref{epsilon_z_t}(b-d), the contour of the reduced invariant electric field is shown in $zt$-plane with CEP $\tilde{\varphi} =\pi/2,3\pi/4,\:\:\textnormal{and}\:\:\:\pi$. It shows that the peak positions are getting shifted in temporal axis with CEP whereas in the $z$-axis, no changes have been observed in the peak positions which is obvious from the simplified expression of $\epsilon^e$ in Eq.\ref{epsilon1_e}. Here due to the shift in the peak positions, the maximum peak height also gets reduced because of the Gaussian pulse envelope function. It ensures that the control over the CEP is important is important in the context of the processes which depend on the peak field strength.

It shows that the maximum electric field is located at $z = 0$, and due to the interference of counterpropagating beams in $z,\: t$ the reduced electric field is distributed like localized spikes in $zt$ -plane. The temporal location of the peak positions is very sensitive to CEP.  For $\tilde{\varphi} =\pi/2$, the contour of the reduced field $\epsilon^e$ is shown in the Fig.\ref{epsilon_z_t}(b) which describes the locations of the maximum field intensities in the $zt$-plane. Here at the central position ($z = t = 0$) the field distribution possesses maximum intensity. Fig.\ref{epsilon_z_t}(c) shows the same for $\tilde{\varphi} = 3\pi/4$ where it reflects the shift in the temporal axis. Significant changes have observed in the contour of the reduced electric field distribution for $\tilde{\varphi} = \pi$ in which the peak field intensity in the central position is zero and it gets shifted in the time axis. 

From the analytical expression of the simplified reduced electric field $\epsilon^e$, we discuss the locations of the peak positions and the corresponding shifts with CEP as presented in Fig.\ref{epsilon_z_t}(a). We consider at the focus ($z = 0$). 
From the Eq.\ref{epsilon1_e}, we have the location of the central peak position as $\omega t+\tilde{\varphi} = \pm \pi/2$, which ends up with two values such as $\omega t_{+} = \pi/2-\tilde{\varphi}$ and $\omega t_{-} = -\pi/2-\tilde{\varphi}$. Here $\pm$ in the subscript denote the temporal positions corresponding positive and negative time axis. So the difference between the locations of the central peaks in positive and negative time axis is given by $\omega (t_{+}-t_{-}) = \pi$ or $(t_{+}-t_{-}) = \pi/\omega$. It concludes that the separation between temporal positions in central peaks are independent on the values of CEP. Some special cases are as follow: (1) For $\tilde{\varphi} = \pi/2$, we have central peak at $\omega t_{central} = 0$ along with two side peaks at $\omega t_+ = \pi$ and $\omega t_- = -\pi$. (2) For $\tilde{\varphi} = \pi/4$,  we have  $\omega t_+ = \pi/4$ and $\omega t_- = -3\pi/4$. (3) $\tilde{\varphi} = 0$, we have $\omega t_+ = \pi/2$ and $\omega t_- = -\pi/2$. So the above analysis and the distribution of reduced electric field in the Fig.\ref{epsilon_z_t} coincide and it tells that the central maxima are changing, depending on the values of CEP. Such features will also contribute in the relative shift in the location of the particle distribution in time which we will discuss in the next section. 
\begin{figure}
\begin{center}
\subfloat[]{\includegraphics[width = 2.4in]{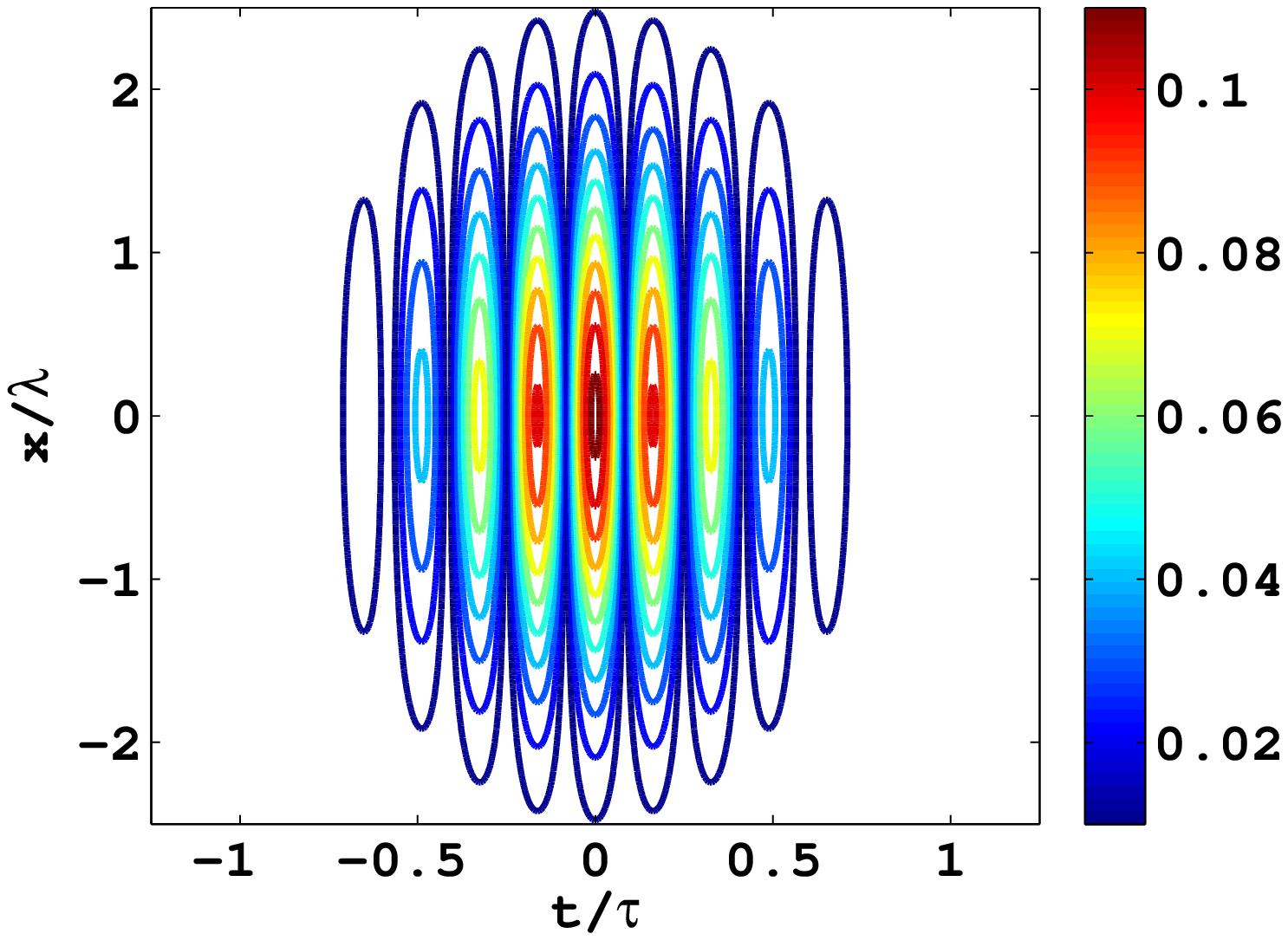}} 
\subfloat[]{\includegraphics[width = 2.4in]{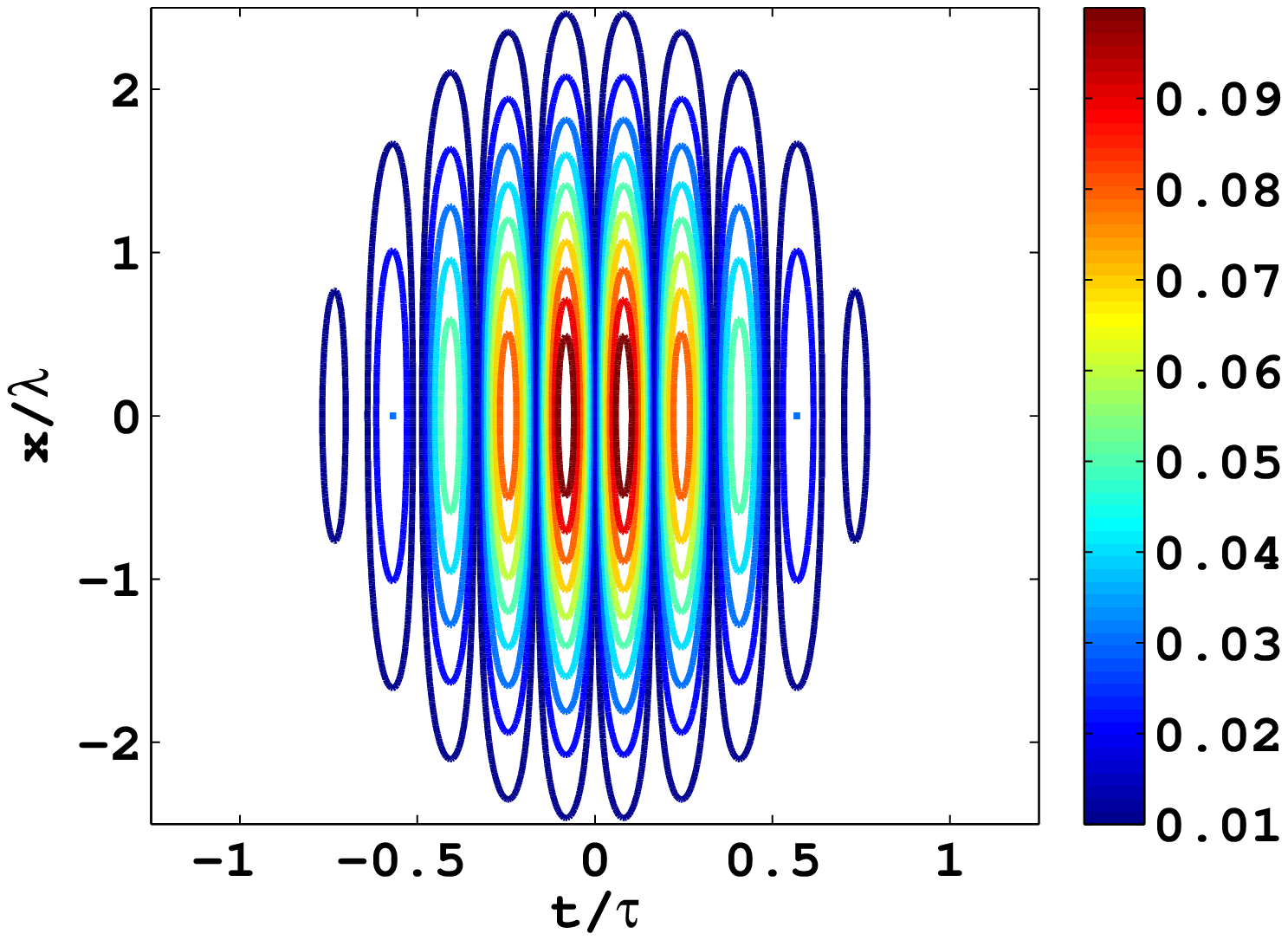}}\\    
\subfloat[]{\includegraphics[width = 2.4in]{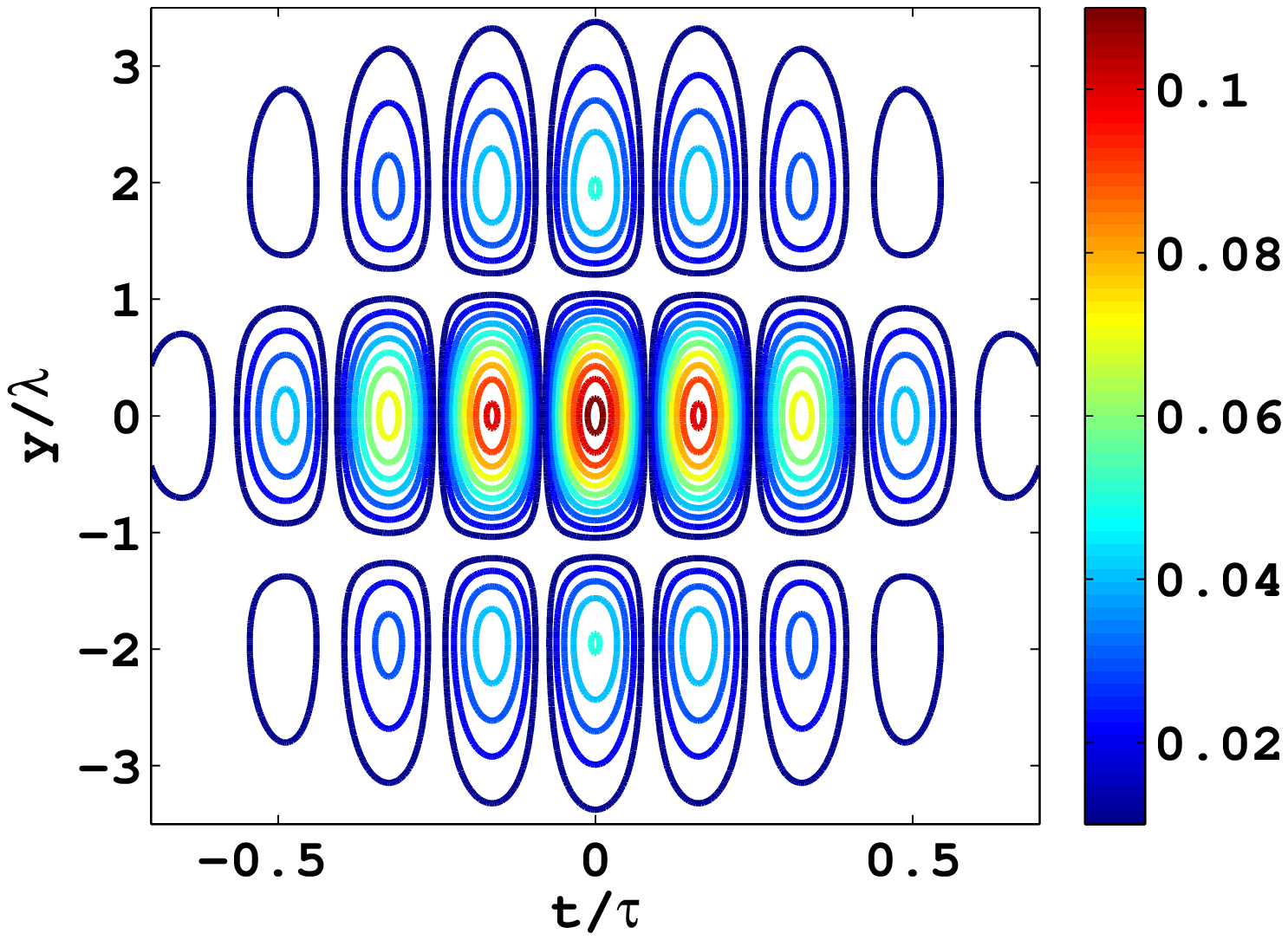}}
\subfloat[]{\includegraphics[width = 2.4in]{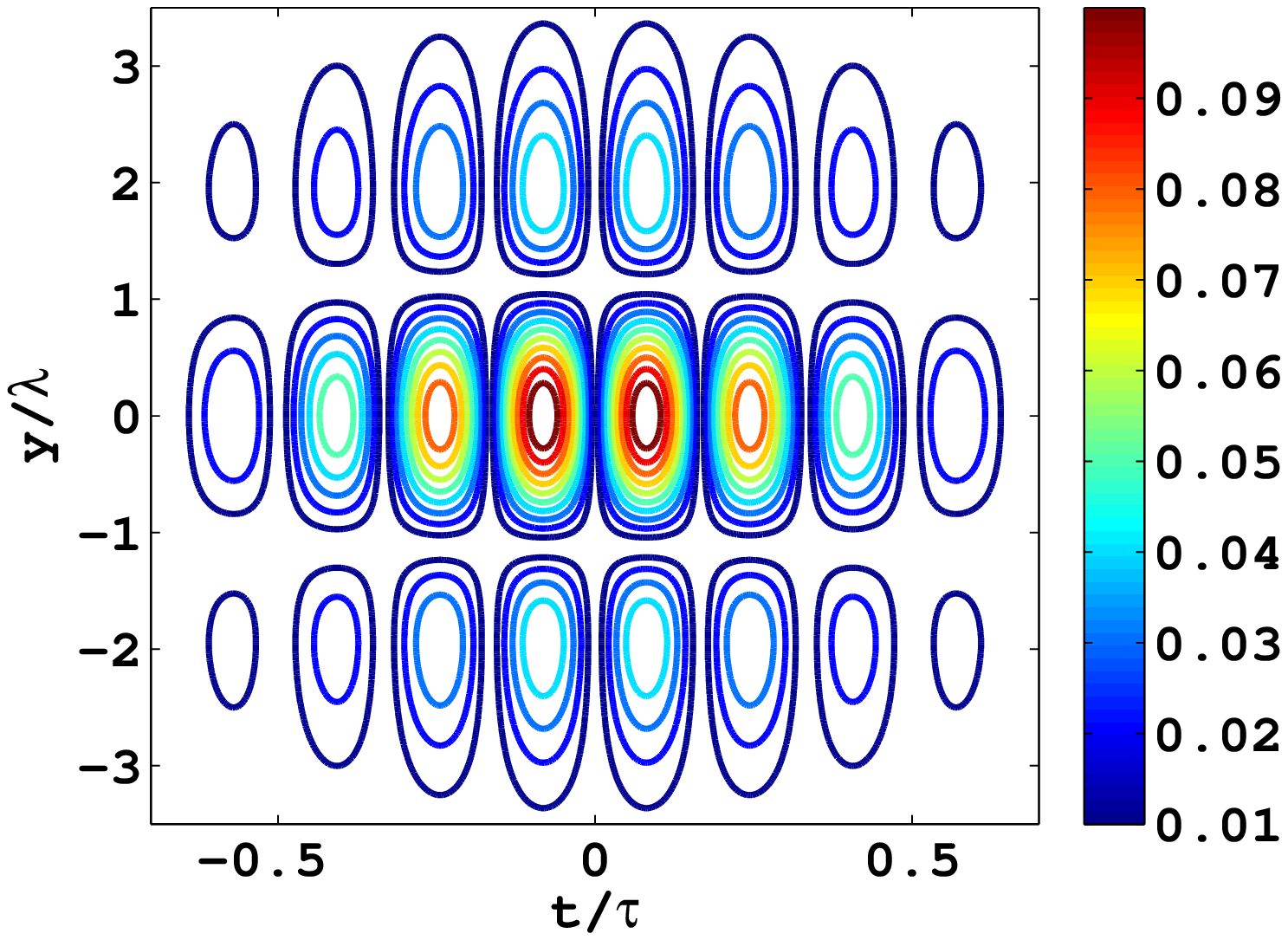}} 
\caption{Contour plots of the $\epsilon^e$ in $xt$-and $yt$-planes in scaled variables for CEP $\tilde{\varphi} = \pi/2$ and $\pi$, showing the locations of the peak field positions. Top left panel shows for $\tilde{\varphi} = \pi/2$ and in the top right it is for $\tilde{\varphi} = \pi$ in the $xt$-plane for $y = z =0$. In the bottom it shows same in the $yt$-plane for $x = z = 0$ are given. The other field parameters are $E_0 = 0.0565$, $\Delta = 0.1$, $\tau = 10fs$, and $\lambda = 1\mu m$. The adjacent colour bars are showing the normalized field strength at the field peak positions.} 
\label{epsilon_x_y_t}
\end{center}
\end{figure}

Fig.\ref{epsilon_x_y_t}(a-b) shows the invariant electric field distributions in $xt$-plane for CEP = $\pi/2$ and $\pi$. It forms oscillatory pattern in time with decreasing amplitude whereas in $x$-axis it forms Gaussian distribution. So we have overall Gaussian waves which are localized in time axis.
But in the $yt$-distribution it presents slightly different picture. In the Fig.\ref{epsilon_x_y_t}(c-d), the invariant electric field distributions in the $yt$-plane has been shown for the same of values of CEP as in the $xt$-distribution. The temporal distributions are same but in $y$ axis, it shows two extra peaks apart from the central maxima which is also obvious from the Eq.\ref{epsilon1_e}. All such invariant electric field distributions control the rate of the particle distributions with space-time coordinates for different values of CEP which we see in Sec.\ref{part_dist}.  
\subsection{Particle distribution}\label{part_dist}
\begin{figure}
\begin{center}
\subfloat[]{\includegraphics[width = 2.4in]{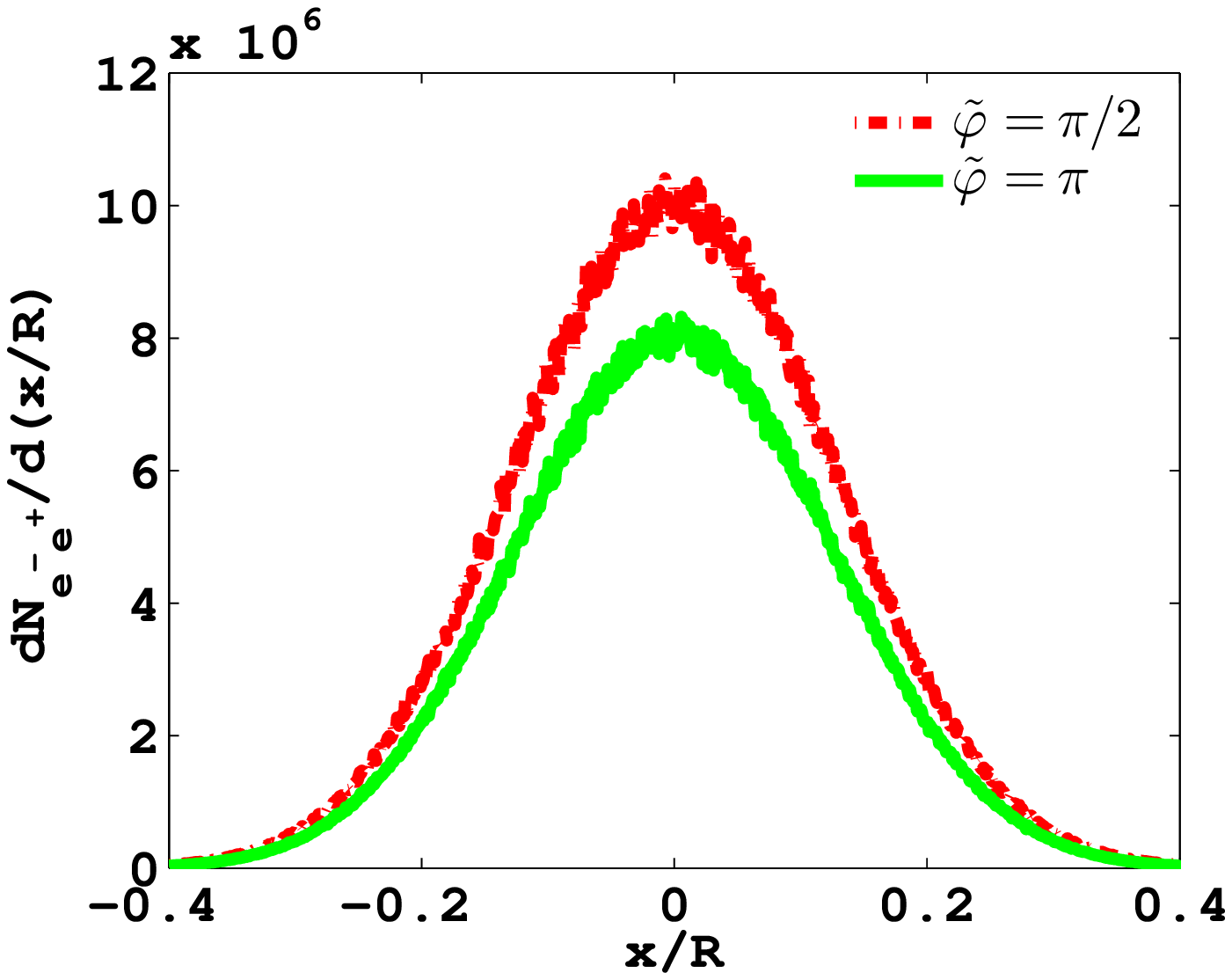}} 
\subfloat[]{\includegraphics[width = 2.4in]{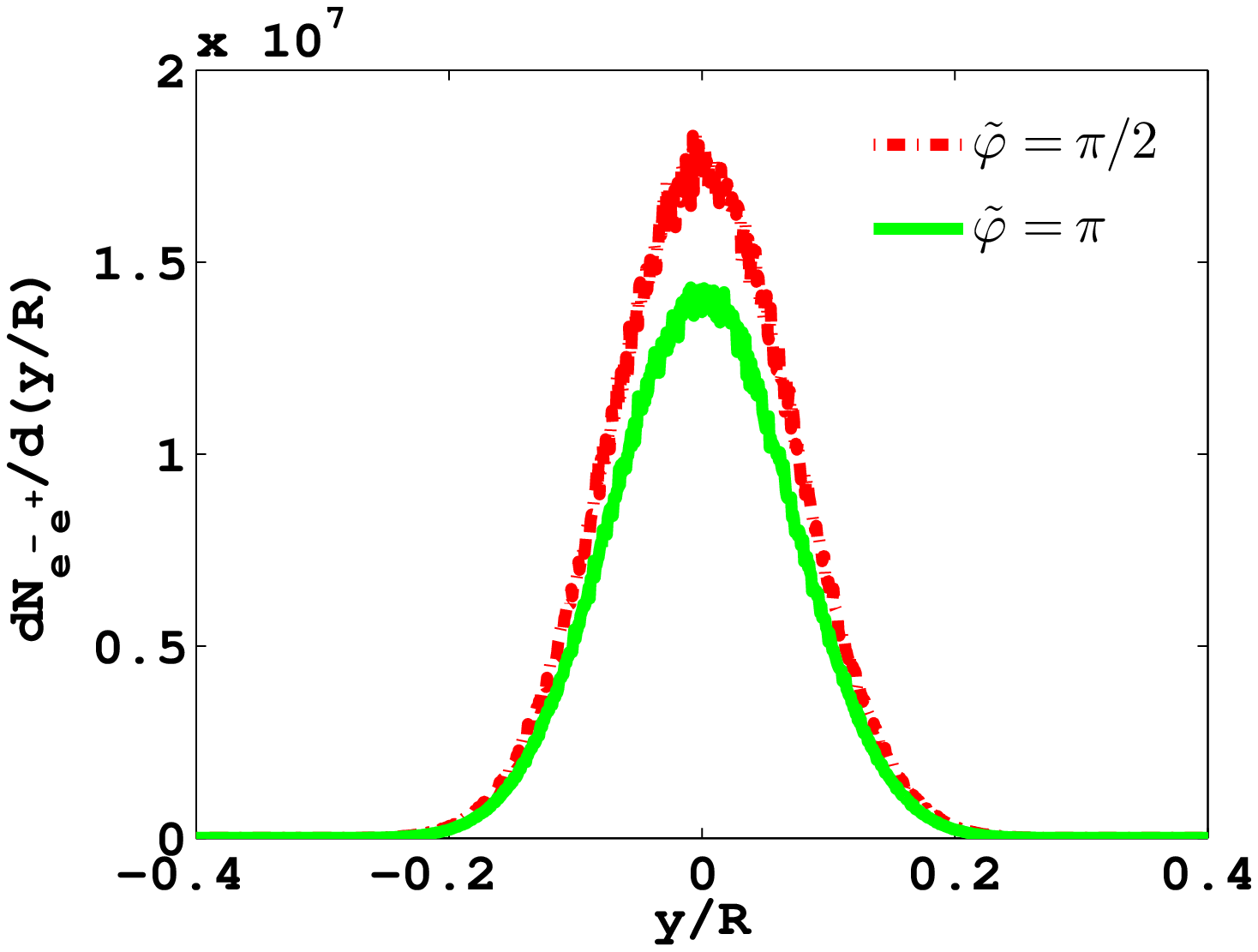}}\\ 
\subfloat[]{\includegraphics[width = 2.4in]{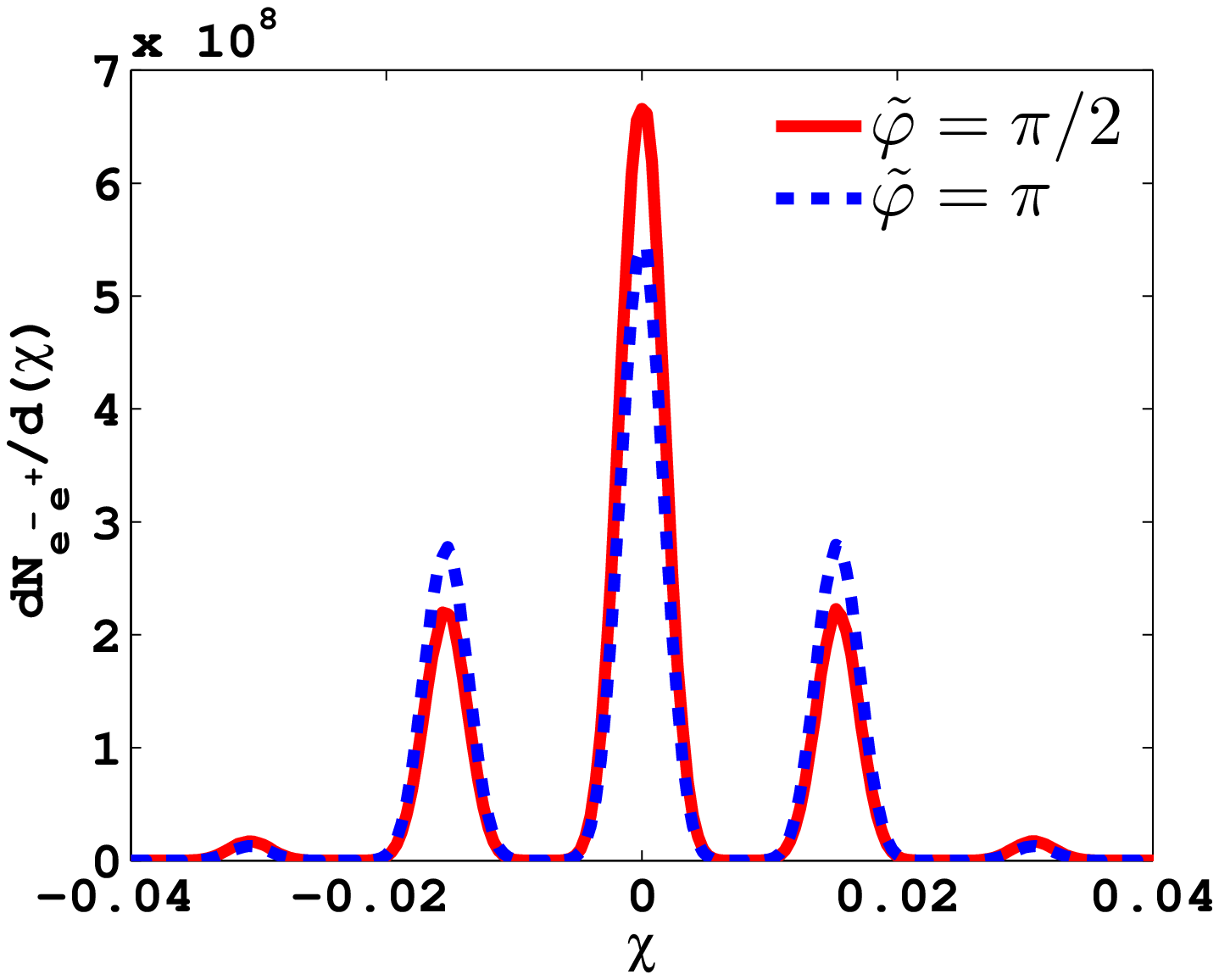}}
\subfloat[]{\includegraphics[width = 2.4in]{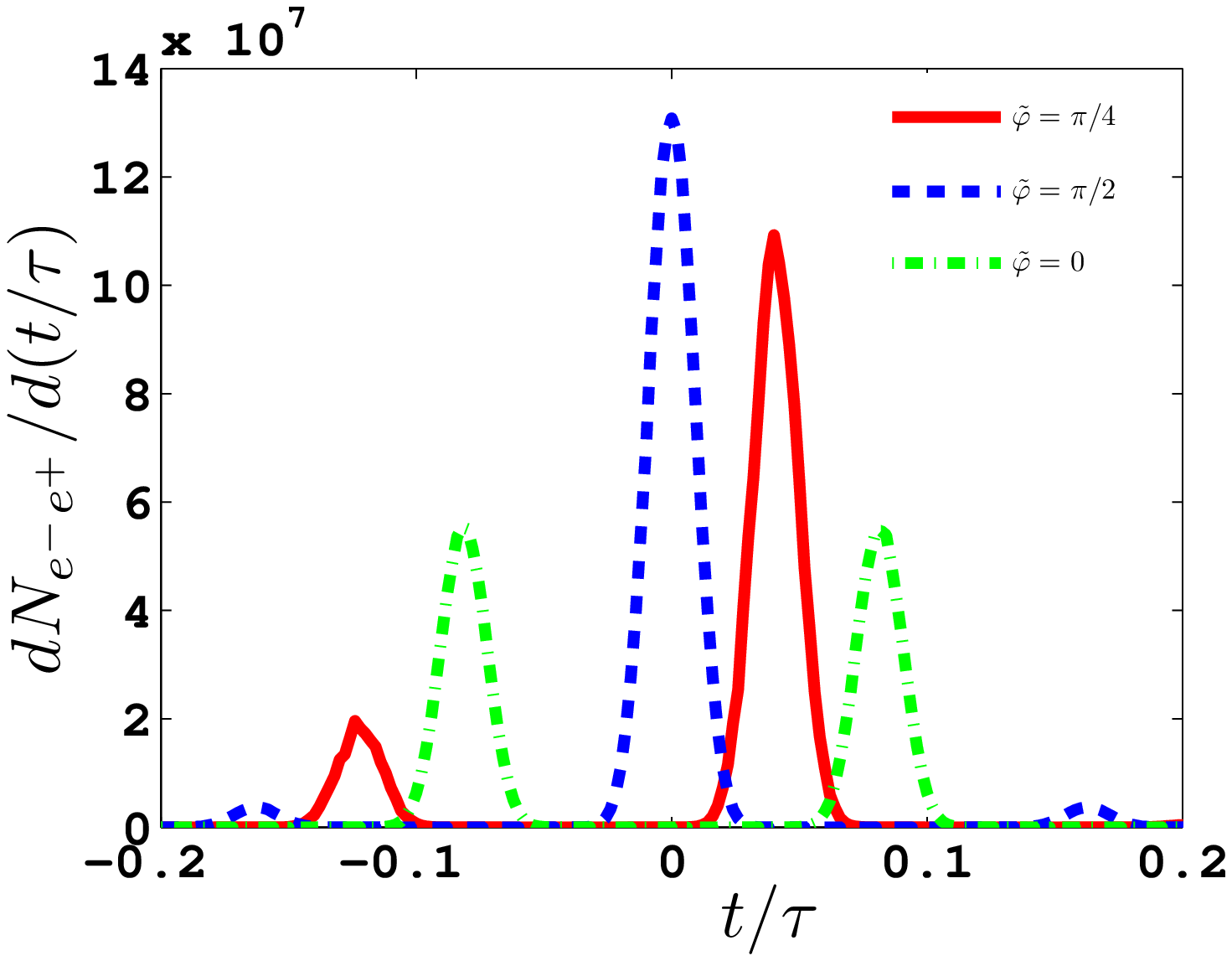}} 
\caption{The differential particle rates in scaled space-time variables for different values of CEP. In the upper channel it shows in the transverse coordinates $x$ and $y$ and in the lower channel it is for longitudinal variable $\chi$ and time. The other field parameters are $E_0 = 0.0565$, $\Delta = 0.1$, $\tau = 10fs$, and $\lambda = 1\mu m$.}
\label{N_x_y_chi_t}
\end{center}
\end{figure}

Here we show the differential pair distribution in space-time coordinates by applying Schwinger formula \cite{PhysRev.82.664} for the average number of particle generation per unit volume and per unit time as given below:
\begin{equation}
 \label{w_e_average}
 w_{e^- e^+} = \frac{dN_{e^- e^+}}{dVdt} = \frac{e^2 E_S^2}{4 \pi^2 \hbar^2 c}  \epsilon^e  \eta^e \coth(\frac{\pi \eta^e}{\epsilon^e})\exp(-\frac{\pi}{\epsilon^e}).
 \end{equation}
Using Eq.\ref{w_e_average}, we calculate the differential pairs numerically by integrating the other coordinates and we show the density of the pair distribution in space-time coordinates such as $x$, $y$, $\chi = z/L$, and in $t$ for e-LPCLP mode for different values of CEP.
First we discuss the $x$-distribution of the differential pairs for two values of CEP ($\pi/2$ and $\pi$) which are the two optimum values for producing maximum and minimum rates and numbers of pairs generation. Fig.\ref{N_x_y_chi_t}(a), it depicts the differential pairs distributions in $x/R$ for CEP $\tilde{\varphi} =\pi/2$ and $\pi$. It forms like a Gaussian profile which is obvious as the reduced electric field distribution exhibits such profiles (Fig.\ref{epsilon_x_y_t}(a-b)). The contour plot of $\epsilon^e$ in normalized $xt$-plane shows Gaussian nature along the $x$ axis and oscillatory nature in the time axis. Because of the extended electric field distribution in the $x$ axis the rate of the pair production gets quite broader.\\
The $dN_{e^+e^-}/d(y/R)$ with $y/R$ is shown in the Fig.\ref{N_x_y_chi_t}(b). It shows a Gaussian profile and CEP $\tilde{\varphi} = \pi/2$ leads to the maximum rate of the differential pair generation in $y$. Such profile can be explained from the contour plot of reduced invariant electric field $\epsilon^e$ as shown in Fig.\ref{epsilon_x_y_t}(c). Here apart from the Gaussian form function in the analytical expression of $\epsilon^e$ in Eq.\ref{epsilon1_e}, it also varies quadratically in $y$. It makes two nodes in $y$ distribution and we have one central maxima and two side peaks. Such field profiles are quite sharp and leads to maximum rate of the pair compared to the $x$ distribution. Because of that the $y$ distribution is localized in $y$-axis and leads to a quite sharp pair distribution. So we have non-identical particle distribution in $x$ and $y$ axes. 
Fig.\ref{N_x_y_chi_t}(c) shows the variation $dN_{e^+e^-}/d\chi$ with $\chi$ which exhibits some interesting features with CEP. We see that the distribution shows one central peak along with two side peaks. For the central peak the maximum value is at $\tilde{\varphi} = \pi/2$, however, for the side peaks, the maximum value is observed at $\tilde{\varphi} = \pi$, rather than $\pi/2$. As the electric field distribution in Fig.\ref{epsilon_z_t}(d), it shows two maxima at the side peak whereas in the Fig.\ref{epsilon_z_t}(a) there is only one side peak. So the pair distribution in the side peak regions are maximum for $\tilde{\varphi} = \pi$. The spatial distribution of the differential particle with normalized longitudinal coordinate shows very spiky nature along the propagation axis. It exhibits more sharp distribution with $\chi$ in comparison to the CPLP case \cite{2016arXiv160601367B}.\\
In Fig.\ref{N_x_y_chi_t}(d) the differential particle distribution with time is shown for CEP  $\tilde{\varphi} = 0,\; \pi/4\:$, and $\pi/2 $. It tells about the sensitivity of the differential pair distribution with CEP. The maximum value of the pair occurs at $\tilde{\varphi} = \pi/2$ and minimum at $0$. The peaks of the pair distributions are being reduced for different values of $\tilde{\varphi}$. 
 Here the shift in the central peak results the reduction of the peak height due to the pulse envelope function. So the temporal distribution of the pairs gets reduced for CEP other than $\pi/2 $. We have observed such asymmetrical distribution for CEP = $\pi/4$. This can be explained by calculating the locations of the reduced electric field maxima. In the central zone we have two points which are located at $t_+ = \pi/{4\omega}$ and $t_- = -3\pi/{3\omega}$. These two values correspond to the reduction of the peak electric field strength differently which causes an asymmetric particle distribution. Because of the CEP, the internal field oscillation advances towards the leading edge of the pulse envelope and peak position of the field is being shifted. So at such positions due to the pulse envelope function due to the reduction in the peak field strength, the temporal rate of the pair generation gets reduced. $\tilde{\varphi} = \pi/2$ corresponds to the central position at which $\epsilon^e$ has the maximum value which basically generates maximum numbers of pairs.\\
The $dN_{e^+e^-}/d(t/\tau)$ with $t/\tau$  is very sharp and it has a FWHM of $200as$ as for a laser pulse duration of $10fs$. Such kind of sharp bunch generation is important for the generation of $e^-$ or $e^+$ beams having small temporal spread, high $\gamma$ value etc. by applying a suitable magnetic field. 
All the differential particle distributions resemble the reduced invariant electric field pattern but its nonlinear dependence on the reduced electric field strength makes the variation more sharp.
\begin{table}
 \caption{$N_{e^+e^-}$ for $\tilde{\varphi}$. Here $E_0 = 0.0565$, $\Delta = 0.1$,\\ $\tau = 10fs$, and $\lambda = 1\mu m$.} 
 \centering  
 \begin{tabular}{c c} 
 \hline\hline                        
  $\tilde{\varphi}$ & $N_{e^+e^-}$  \\ [2ex] 
 \hline                  
 $0$ & $2.517\times 10^6$ \\ 
 \hline
 $\frac{\pi}{8}$ & $2.605\times 10^6$ \\
 \hline
 $\frac{\pi}{4}$ & $2.856\times 10^6$  \\
  \hline
$\frac{3\pi}{8}$ & $3.051\times 10^6$ \\
 \hline
  $\frac{\pi}{2}$ & $3.157\times 10^6$  \\
 \hline
  $\frac{3\pi}{4}$ & $2.826\times 10^6$   \\
  \hline
 $\frac{5\pi}{6}$ & $2.651\times 10^6$  \\ [2ex]      
 \hline\hline 
  \end{tabular}
  \label{table:N_vs_CEP} 
  \end{table}
For quantitative estimation, we have calculated the average pair number for different values of CEP as shown in the Table \ref{table:N_vs_CEP}. It shows that for $\tilde{\varphi} = \pi/2$, the particle production is maximum whereas for  $\tilde{\varphi} = 0$, it leads to the minimum number of pair production. The average numbers of created pairs are shown with CEP. We see that the most favourable situation occurs at $\tilde{\varphi} = \pi/2$ where the average number of created pairs are maximum. 
\section{conclusion}\label{concl}
To conclude, we have examined the pair production process via Schwinger mechanism for the linearly polarized focused EM field in counterpropagating configuration. First, we have presented the analytical expressions of the EM field, the EM field invariants, and the invariant electric and magnetic fields in the transformed frame. We have shown that the CEP dependence is very important for the linearly polarized beam configuration in comparison to the circular polarization. We have also shown  that the invariant field distribution depends on the CEP predominantly in time. It exhibits oscillatory behaviour in $z$ and time which causes electric field distribution more spiky and more localized in space time region. Such properties have been reflected in the differential particle distributions in $z$ and time axes. We have seen that the particle distributions in space-time coordinates depends on the CEP. The significant changes have been observed in the temporal distribution of pairs. For CEP = $\pi/2$ we have an ultrashort particle bunch production whose central maxima is located at $t = 0$. However for CEP = $0$ and $\pi/4$ we have two peaks in distribution profiles and the later on produces an asymmetrical distribution of the pairs.
For linearly polarized laser pulses, the invariant field persists extra condition between the dynamical phase and the CEP. It defines two regions by the Lorentz invariants,whether it exhibits electric or magnetic  nature. So for a certain beam configuration, one can choose/control the phase relationship of the focused laser pulses such that it shows electric or magnetic nature in the transformed frame. Hence it gives rise to a mechanism by controlling the phase relationship between instantaneous laser EM wave phase and the CEP to achieve the desired peak field intensity without having energy loss. 
Some of the interesting properties of pair production process such as kinetic nature of the created pairs or the momentum distribution have not been discussed here which will be addressed in forthcoming article.
\begin{acknowledgments}
It is a pleasure to acknowledge the helpful discussions with Mr. M. Kumar and Dr. H. Singhal in the context of CEP dependence in ultrashort laser induced processes.
\end{acknowledgments}
\appendix
\section{CEP dependence on the linear polarization}\label{Appendix A}
For the linearly e-polarized focused EM fields, the expressions of the electric field in both forward (in $z$ direction) and backward (in $-z$ direction) propagations can be written as \cite{bula}
\begin{equation}\label{E_f_single}
\textbf{E}_f^e = iE_0e^{-i\omega(t-z/c)-i\tilde{\varphi}}g\Bigg[\hat{\textbf{e}}_x(F_1-F_2\cos{2\phi})-\hat{\textbf{e}}_yF_2\sin{2\phi}\Bigg],
\end{equation}
and
\begin{equation}\label{E_b_single}
\textbf{E}_b^e = iE_0e^{-i\omega(t+z/c)-i\tilde{\varphi}}g\Bigg[\hat{\textbf{e}}_x(F_1^*-F_2^*\cos{2\phi})-\hat{\textbf{e}}_yF_2^*\sin{2\phi}\Bigg].
\end{equation}
Here $F_1,\: F_2$ are the Gaussian form functions for the focused laser beam \cite{Narozhny2000} given as
 \begin{equation}\label{Form_fuctions}
 \begin{split}
 F_1 = (1+2i\chi)^{-2} ( 1-\frac{\xi^2}{1+2i\chi} ) \exp(-\frac{\xi^2}{1+2i\chi}),\:\:\textnormal{and}\:\: 
 F_2 = -\xi^2(1+2i\chi)^{-3} \exp(-\frac{\xi^2}{1+2i\chi}).
\end{split}
\end{equation}
 $F_1^*,\: F_2^*$ are the complex conjugate of them. All the symbols have already been defined in \ref{field_theory}. Similarly one can have the expressions magnetic field in forward and backward directions as \cite{bula}
\begin{equation}\label{H_f_single}
\textbf{H}_f^e = iE_0e^{-i\omega(t-z/c)-i\tilde{\varphi}}g\Bigg[(1-i\Delta^2\frac{\partial}{\partial\chi})\Big\{\hat{\textbf{e}}_xF_2\sin{2\phi}-\hat{\textbf{e}}_y(F_1-F_2\cos{2\phi}\Big\}+2i\Delta\sin{\phi}\frac{\partial F_1}{\partial\xi}\Bigg],
\end{equation}
and 
\begin{equation}\label{H_b_single}
\textbf{H}_b^e = -iE_0e^{-i\omega(t+z/c)-i\tilde{\varphi}}g\Bigg[(1+i\Delta^2\frac{\partial}{\partial\chi})\Big\{\hat{\textbf{e}}_xF_2^*\sin{2\phi}-\hat{\textbf{e}}_y(F_1^*-F_2^*\cos{2\phi}\Big\}+2i\Delta\sin{\phi}\frac{\partial F_1^*}{\partial\xi}\Bigg].
\end{equation}
Now if we allow them to superimpose in the focal region, the expressions of the electric and magnetic fields are given by
\begin{equation}\label{lin_2_E}
\textbf{E}^e = \textbf{E}^e_f+\textbf{E}^e_b = 2iE_0e^{-i(\omega t+\tilde{\varphi})}g\Bigg[\hat{\textbf{e}}_xRe\Big[(F_1-F_2\cos{2\phi})e^{i\omega z/c}\Big]-\hat{\textbf{e}}_yRe\Big[F_2e^{i\omega z/c}\sin{2\phi}\Big]\Bigg]
\end{equation}
\begin{equation}\label{lin_2_H}
\begin{split}
\textbf{H}^e = \textbf{H}^e_f+\textbf{H}^e_b = -2E_0e^{-i(\omega t+\tilde{\varphi})}g\Bigg[\hat{\textbf{e}}_x Im\Big[F_2e^{i\omega z/c}\sin{2\phi}\Big]+ \hat{\textbf{e}}_y Im\Big[(F_1-F_2\cos{2\phi})e^{i\omega z/c}\Big]\\+2i\Delta\sin{\phi}Im\Big[e^{i\omega z/c}\frac{\partial F_1}{\partial\xi}\Big]\Bigg]
\end{split}
\end{equation}
But the physical electric and magnetic fields are real part of Eqs.(\ref{lin_2_E},\ref{lin_2_H}) which are given as
\begin{equation}\label{lin_2_E_Re}
\begin{split}
Re\textbf{E}^e = 2E_0\sin(\omega t+\tilde{\varphi})g\Bigg[\hat{\textbf{e}}_xRe\Big[(F_1-F_2\cos{2\phi})e^{i\omega z/c}\Big]-\hat{\textbf{e}}_yRe\Big[F_2e^{i\omega z/c}\sin{2\phi}\Big]\Bigg]
 = 2E_0g\frac{e^{-\xi^2/{(1+4\chi^2)}}}{1+4\chi^2}\sin(\omega t+\tilde{\varphi})\\\times\Bigg[\hat{\textbf{e}}_x\Big\{\cos(\omega z/c-2\psi)-\frac{2\xi^2}{(1+4\chi^2)^{1/2}}\sin^2{\phi}\cos(\omega z/c-3\psi)\Big\}+\hat{\textbf{e}}_y\frac{\xi^2}{(1+4\chi^2)^{1/2}}\sin{2\phi}\cos(\omega z/c-3\psi)\Bigg]
\end{split}
\end{equation}
and 
 \begin{equation}\label{lin_2_H_Re}
\begin{split}
Re\textbf{H}^e = -2E_0g\Bigg[\cos(\omega t+\tilde{\varphi})\Big\{\hat{\textbf{e}}_x Im\Big[F_2e^{i\omega z/c}\sin{2\phi}\Big]+ \hat{\textbf{e}}_y Im\Big[(F_1-F_2\cos{2\phi})e^{i\omega z/c}\Big]\Big\}+2\Delta\sin{\phi}\sin(\omega t+\tilde{\varphi})\\\times\Big[e^{i\omega z/c}\frac{\partial F_1}{\partial\xi}\Big]\Bigg]
 \approx -2E_0g\frac{e^{-\xi^2/{(1+4\chi^2)}}}{1+4\chi^2}\Bigg[\cos(\omega t+\tilde{\varphi})\Bigg(\hat{\textbf{e}}_x\frac{\xi^2}{(1+4\chi^2)^{1/2}}\sin{2\phi}\sin(\omega z/c-3\psi)-\hat{\textbf{e}}_y\Big\{\sin(\omega z/c-2\psi)\\-\frac{2\xi^2}{(1+4\chi^2)^{1/2}}\sin^2{\phi}\sin(\omega z/c-3\psi)\Big\}\Bigg)+4\xi\Delta\frac{\sin(\omega t+\tilde{\varphi})}{(1+4\chi^2)^{1/2}}\sin{\phi}\sin(\omega z/c-3\psi)\hat{\textbf{e}}_z\Bigg].
\end{split}
\end{equation}
To derive the approximate Eqs.(\ref{lin_2_E_Re},\ref{lin_2_H_Re}) we have used the expressions of $F_1,\: F_2$ from the Eq.(\ref{Form_fuctions}) in the small $\chi,\:\xi$ limit.
It shows that resultant fields are oscillating in longitudinal coordinate and time also. It also shows the CEP dependence in the leading order term. \\
\textbf{Appendix B: CEP dependence on the circular polarization}\label{Appendix B}\\
Here we discuss CEP dependence on circular polarization in the counterpropagating configuration. The electric and magnetic fields expressions propagating forward direction (in $+z$ direction) focused laser beam having CEP between carrier wave and the envelope function are given as
\begin{equation}
\label{Electric_field_e_single_beam_f}
 \textbf{E}_f^e = iE_0e^{-i\omega(t-z/c)-i\tilde{\varphi}}g\Bigg\{F_1(\hat{\textbf{e}}_x + i\hat{\textbf{e}}_y)-F_2e^{2i\phi}(\hat{\textbf{e}}_x - i\hat{\textbf{e}}_y)\Bigg\},
 \end{equation}
and
\begin{equation}\label{Magnetic_field_e_single_beam_f}
\textbf{H}_f^e = E_0e^{-i\omega(t-z/c)-i\tilde{\varphi}}g\Bigg\{(1-i\Delta^2\frac{\partial}{\partial\chi})\Big[F_1(\hat{\textbf{e}}_x + i\hat{\textbf{e}}_y)+F_2e^{ 2i\phi}(\hat{\textbf{e}}_x - i\hat{\textbf{e}}_y)\Big]+2i\Delta e^{i\phi}\frac{\partial F_1}{\partial\xi}\hat{\textbf{e}}_z\Bigg\}.
\end{equation}
Similar expressions of the EM field in the backward direction (in $-z$ direction) can be written as 
\begin{equation}
\label{Electric_field_e_single_beam_b}
 \textbf{E}_b^e = iE_0e^{-i\omega(t+z/c)-i\tilde{\varphi}}g\Bigg\{F_1^*(\hat{\textbf{e}}_x + i\hat{\textbf{e}}_y)-F_2^*e^{-2i\phi}(\hat{\textbf{e}}_x - i\hat{\textbf{e}}_y)\Bigg\},
  \end{equation}
and
 \begin{equation}\label{Magnetic_field_e_single_beam_b}
\textbf{H}_b^e = - E_0e^{-i\omega(t+z/c)-i\tilde{\varphi}}g\Bigg\{(1+i\Delta^2\frac{\partial}{\partial\chi})\Big[F_1^*(\hat{\textbf{e}}_x + i\hat{\textbf{e}}_y)+F_2^*e^{ -2i\phi}(\hat{\textbf{e}}_x - i\hat{\textbf{e}}_y)\Big]+2i\Delta e^{-i\phi}\frac{\partial F_1^*}{\partial\xi}\hat{\textbf{e}}_z\Bigg\}.
\end{equation}
At the focus the resultant EM field structures due to the superposition of forward and backward beams are
\begin{equation}\label{cir_E_two}
\begin{split}
\textbf{E}^e = \textbf{E}_f^e+\textbf{E}_b^e = 2iE_0e^{-i(\omega t+\tilde{\varphi})}g\Bigg\{(\hat{\textbf{e}}_x + i\hat{\textbf{e}}_y)Re\Big[F_1e^{i\omega z/c}\Big]-(\hat{\textbf{e}}_x - i\hat{\textbf{e}}_y)Re\Big[F_2e^{2i\phi}e^{i\omega z/c}\Big]\Bigg\},
\end{split} 
\end{equation}
and
\begin{equation}\label{cir_H_two}
\begin{split}
\textbf{H}^e = \textbf{H}_f^e+\textbf{H}_b^e = 2iE_0e^{-i(\omega t+\tilde{\varphi})}g\Bigg\{(\hat{\textbf{e}}_x + i\hat{\textbf{e}}_y)Im\Big[F_1e^{i\omega z/c}\Big]+(\hat{\textbf{e}}_x - i\hat{\textbf{e}}_y)Im\Big[F_2e^{2i\phi}e^{i\omega z/c}\Big]\\+2i\Delta Im\Big[e^{i\phi}e^{i\omega z/c}\frac{\partial F_1}{\partial\xi}\Big]\hat{\textbf{e}}_z\Bigg\}.
\end{split} 
\end{equation}
The physical electric and magnetic fields are real part of the Eqs.(\ref{cir_E_two},\ref{cir_H_two}) which are given as
\begin{equation}\label{cir_E_two_Re}
Re\textbf{E}^e = 2E_0g\Bigg[\sin(\omega t+\tilde{\varphi})Re\Big[(F_1-F_2e^{2i\phi})e^{i\omega z/c}\hat{\textbf{e}}_x-\cos(\omega t+\tilde{\varphi})Re\Big[(F_1+F_2e^{2i\phi})e^{i\omega z/c}\hat{\textbf{e}}_y\Big]\Bigg],
\end{equation}
and
\begin{equation}\label{cir_H_two_Re}
\begin{split}
Re\textbf{H}^e = 2E_0g\Bigg[\sin(\omega t+\tilde{\varphi})Im\Big[(F_1+F_2e^{2i\phi})e^{i\omega z/c}\hat{\textbf{e}}_x-\cos(\omega t+\tilde{\varphi})Im\Big[(F_1-F_2e^{2i\phi})e^{i\omega z/c}\hat{\textbf{e}}_y\Big]\\-2\Delta\cos(\omega t+\tilde{\varphi})Im\Big[e^{i\phi}e^{i\omega z/c}\frac{\partial F_1}{\partial\xi}\Big]\hat{\textbf{e}}_z\Big]\Bigg].
\end{split}
\end{equation}
Now using the expressions of $F_1$ and $F_2$ Eq.(\ref{Form_fuctions}), we have the expressions of the electric and magnetic fields as
 \begin{equation}\label{cir_E_two_Re_approx}
 \begin{split}
 Re\textbf{E}^e = 2E_0g\frac{e^{-\frac{\xi^2}{{1+4\chi^2}}}}{(1+4\chi^2)}\Bigg[\sin(\omega t+\tilde{\varphi})\Big\{\cos\left(\omega z/c-2\psi\right)-\frac{2\xi^2\sin{\phi}}{(1+4\chi^2)^{1/2}}\sin\left(\phi+\omega z/c-3\psi\right)\Big\}\hat{\textbf{e}}_x\\-\cos(\omega t+\tilde{\varphi})\Big\{\cos\left(\omega z/c-2\psi\right)-\frac{2\xi^2\cos{\phi}}{(1+4\chi^2)^{1/2}}\cos(\phi+\omega z/c-3\psi)\Big\}  \hat{\textbf{e}}_y\Bigg],
 \end{split}
 \end{equation}
and
\begin{equation}\label{cir_H_two_Re_approx}
 \begin{split}
 Re\textbf{H}^e = 2E_0g\frac{e^{-\frac{\xi^2}{{1+4\chi^2}}}}{(1+4\chi^2)}\Bigg[\sin(\omega t+\tilde{\varphi})\Big\{\sin\left(\omega z/c-2\psi\right)-\frac{2\xi^2\cos{\phi}}{(1+4\chi^2)^{1/2}}\sin\left(\phi+\omega z/c-3\psi\right)\Big\}\hat{\textbf{e}}_x\\-\cos(\omega t+\tilde{\varphi})\Big\{\sin\left(\omega z/c-2\psi\right)-\frac{2\xi^2\sin{\phi}}{(1+4\chi^2)^{1/2}}\cos(\phi+\omega z/c-3\psi)\Big\}  \hat{\textbf{e}}_y\\-\frac{8\Delta\xi}{(1+4\chi^2)^{1/2}}(1-\frac{\xi^2}{2(1+4\chi^2)^{1/2}})\cos(\phi+\omega z/{c})\cos(\omega t+\tilde{\varphi})\hat{\textbf{e}}_z\Bigg].
 \end{split}
 \end{equation}
So at the field magnitude level of EM fields are given by 
\begin{equation}\label{ReE_mu_1}
\begin{split}
 Re|\textbf{E}^e| \approx \frac{2E_0 ge^{-\frac{\xi^2}{1+4\chi^2}}}{(1+4\chi^2)}|\cos(\omega z/c-2\psi)|\Bigg[1-\frac{\xi^2}{\cos(\omega z/c-2\psi)(1+4\chi^2)^{1/2}}\Big\{\cos(\omega z/c-3\psi)\\+\cos{2(\omega t+\tilde{\varphi})}\cos(3\psi-\omega z/c-2\phi)\Big\}+\mathcal{O}(\xi^4)\Bigg],
\end{split}
\end{equation}
 and
\begin{equation}\label{ReH_mu_1}
\begin{split}
Re|\textbf{H}^e| \approx \frac{2E_0 ge^{-\frac{\xi^2}{1+4\chi^2}}}{(1+4\chi^2)}|\sin(\omega z/c-2\psi)|\Bigg[1-\frac{\xi^2}{\sin(\omega z/c-2\psi)(1+4\chi^2)^{1/2}}\Big\{\sin(\omega z/c-3\psi)\\+\cos{2(\omega t+\tilde{\varphi})}\sin(3\psi-\omega z/c-2\phi)\Big\}+\mathcal{O}(\xi^4)\Bigg].
\end{split}
\end{equation}
The above analysis shows that for the circular polarization the CEP dependence exists in the individual field components in the lab frame. However at the field magnitude level in the leading order term near the focus, the expressions in Eqs.(\ref{ReE_mu_1},\ref{ReH_mu_1}) are free from fast oscillation and it is also independent of CEP. It can easily be shown that all the invariant electric and magnetic fields will be independent of CEP. This observation is obvious because for circular polarization, the electric field vector is circulating and any constant phase (here the phase difference between the carrier wave and the envelope function) does not cause any significant change at the field magnitude level and hereafter. 
\bibliographystyle{h-elsevier} 
\bibliography{ref} 
\end{document}